\documentclass[]{svjour3}
\usepackage{times}
\usepackage{amsmath}
\usepackage{amssymb}
\usepackage{graphicx}

\begin{document}

\title{J\"{u}rgen Ehlers - and the Fate of the Black-Hole Spacetimes}
\author{Wolfgang Kundt}
\institute{W. Kundt \at Argelander Institut f\"{u}r Astronomie der Universit\"{a}t Bonn}
\maketitle
\begin{abstract}
This article -- written in honour of J\"{u}rgen Ehlers -- consists of two
different, though interlocking parts: Section 1 describes my 54 years of
perpetual experiences and exchanges with him, both science and episodes, whilst
section 2 describes the history of astrophysical black holes, which evolved
during the same epoque though largely independently, with its activity centers
in other places of the globe, and has by no means terminated.
\end{abstract}

\section{ J\"urgen Ehlers: our 54 years of close Interaction}

\label{sec:intro}J\"{u}rgen Ehlers and I both studied physics and mathematics
at Hamburg University. My studies began with the summer\ `Semester' of 1950,
J\"{u}rgen's one year earlier; (he was born on 29 November 1929, $\gtrsim$ 18
months before me). When we scientifically matured to terminate our standard
education -- in the fall of 1953, I think -- we were both attracted by the
personality of Pascual Jordan, and jointly started weekly seminar work with
him. Soon we were joined by Engelbert Sch\"{u}cking, and soon again, Pascual
Jordan guided our collaboration, both mentally, and financially. The Hamburg
group in General Relativity had been founded.

J\"{u}rgen, Engelbert, myself, and `der Meister' took turns as weekly seminar
speakers. Initially, Jordan's contributions aimed at improving the second
edition of his book `Schwerkraft und Weltall'. J\"{u}rgen's contributions
educated us primarily on other people's work: they helped us understand it
thoroughly. And Engelbert excited us with brand new stuff which had somehow
entered his mind; unless he did not show up at all, whilst absorbing this
stuff, possibly in a bar near the train station. My own contributions were
often somewhat premature, introducing new problems to the group in order to
receive feedback, and to profit particularly from J\"{u}rgen's suggestions for
deepened understanding. On 4 June 1955, J\"{u}rgen met (his later wife) Anita,
on a large faculty dance in the (decorated) mensa of the university which I
had organised.

Our scientific work was done largely independently, mainly at home, except
when an `airforce report' had to be written, for Jordan's contract with the
U.S. army, or when during the early 60s, survey papers had to be composed for
the `Akademie der Wissenschaften und der Literatur in Mainz', of which Jordan
was a member. Such joint publications helped us gain a certain group
knowledge. But there was enough time left for continued learning, and for
approaching new problems. We tried to get deeper insights into the different
known classes of exact solutions of Einstein's field equations, their
invariant properties, and global continuations. Besides, J\"{u}rgen improved
his -- and our -- understanding of relativistic thermodynamics,
electromagnetism within GR, and relativistic statistical mechanics. And we
searched for reliable ways to decide when two given line elements could be
transformed into each other, by a suitable change of coordinates.

This program began with a discussion of Kurt Goedel's (1949) homogeneous and
stationary cosmological model (with a non-vanishing cosmological constant
$\Lambda$), in which suitably accelerated observers can travel into their own
past. In 1957, our work received a forceful boost by Felix Pirani's physical
evaluation of A.S. Petrov's 1954 algebraic classification of the Weyl tensor
for normal-hyperbolic metrical spaces, i.e. of the traceless Riemann tensor,
as for vacuum spacetimes. Petrov's classification turned out to be a valuable
tool for analytic evaluations, in particular of (strong) gravitational-wave
fields, but also for other classes of\ `algebraically degenerate' spacetimes
in which the eigen values of the Weyl tensor tend to yield invariant
coordinates. Other tools for characterisation came from Elie Cartan's (1946)
basic work on Riemannian spacetimes, in particular a solution of the (local)
`equivalence problem' -- \ later also called\ `isometry problem' \ -- \ of two
given spacetimes, and from results on algebraic invariants by Weitzenb\"{o}ck
(1923), Eisenhart (1927, 1933, 1935, 1949), Thomas (1934), Weyl (1939),
Chevalley (1946), and Ruse (1946), as well as by Blaschke (1950), Taub (1951),
Schouten (1954), Boerner (1955), Komar (1955), Lichnerowicz (1955), Yano
(1955), and Trautman (1958). We thus arrived in addition at the `theorem of
complementarity' (of scalar differential invariants and group trajectories)
which states that a possible r-dim group of isometries of a given $n$-dim
spacetime V$_{n}$ -- with q-dim trajectories, q := r-s -- \ contains
complementary information about its geometry in the sense that one can find
$n$-q independent scalar invariants which complete a system of q symmetry
coordinates to a full (local) $n$-dim coordinate system of V$_{n}$ ; where s
= r-q denotes the dimension of the isotropy group of a generic point.
Homogeneous subspaces are invariantly embedded.

These two theorems -- on constructing (or not) an isometry between two given
V$_{n}$ with or without homogeneous subspaces -- were contained in my 1958
ph.d. thesis (including proof sketches), and can be found again in my 1960 and
1962 joint publications with J\"{u}rgen listed in the references, though only
partially highlighted as theorems. J\"{u}rgen had done the proof-reading at
Hamburg whilst I was in Syracuse (N.Y.) -- working in Peter Bergmann's group
for a bit over a year -- at an epoch with only snail-mail correspondence.
J\"{u}rgen knew that I had discovered an oversight (in my first writeup, and
in other people's proofs) which applies only in the rare case of spaces with
an indefinite metric, and of a special type of their Riemann tensor, for which
the complete system of rational scalar invariants (formed from their metric
and Riemann tensor and its covariant derivatives) gets functionally dependent,
so that one has to take recourse to new methods of constructing scalar
invariants, involving differential calculus. One such nasty special case
turned up in the subclass of the pp waves (see below), with higher (than the usual
2-dim) symmetry, and served as a counter example to published results even
though it was fairly clear how to repair them. Should we have formulated the
theorems as conjectures, or should we have relegated them to the text?
J\"{u}rgen's honest mind preferred the latter.

The years 1959-1962 were brainstormy ones for us. The Petrov-Pirani insight
had likewise reached `relativists' like Ivor Robinson, who had shown me how to
integrate the gravitational field equations analytically, in suitable
coordinate gauges, and Andrzej Trautman, Hermann Bondi, Ray Sachs, Roger
Penrose, and Ted Newman, among others, who showed us\ how to analyse the
far-field behaviour of radiating sources. For me in particular waited the
class of plane-fronted gravitational waves, plus its subclass with parallel
rays -- called `pp waves' for short -- which turned out to be extremely
instructive: The plane-fronted gravitational waves show strong resemblances to
their electromagnetic analogs in flat spacetime, but are richer as a class.
They are among Einstein's vacuum fields with the largest groups of isometries,
and with the most extreme behaviour of their scalar invariants, as already
mentioned above. Their action on clouds of test particles could be elucidated,
and they could be shown to be geodetically complete. Clearly, during the years
of their analysis, J\"{u}rgen's interest and advice meant a great challenge to me.

But like every human activity, Hamburg's relativity group did not enjoy
eternal life: The year 1962 saw me build my house in Hamburg-Hoheneichen, and
move into it in November that same year. J\"{u}rgen and Anita spent several
years abroad, between 1961 and 1971, in Syracuse, Dallas, and Austin (Texas),
and Ulrike and I spent our honeymoon year 1966/7 in Pittsburgh (Pa), as guests
of Ted Newman and his family. Pascual Jordan's 65th anniversary was celebrated
(in Hamburg) in October 1967. In all these years, our relativity group was
kept alive by new (younger) members, like Manfred Tr\"{u}mper, Bernd Schmidt,
Wolf Beiglb\"{o}ck, Klaus Bichteler, Volker En\ss, Michael Streubel, Eva
Ruhnau, Hans-J\"{u}rgen Seifert, Henning Mueller zum Hagen, Hans Seiler, and a
few more. We met at international conferences, like the Texas Symposia on
Relativistic Astrophysics (introduced by Ivor Robinson, Alfred Schild, and
Engelbert Sch\"{u}cking), the Varenna Schools (on lake Como), the Royal
Society Meetings in London, and the Marcel Grossmann Meetings (organised by
Remo Ruffini, beginning at Trieste). There were frequent, lively international cooperations.

There was another intense collaboration during the early 60s, of J\"{u}rgen
and myself: we partcipated actively in a weekly seminar on the selected book
on QED by Jauch \& Rohrlich, jointly with Hans-J\"{u}rgen Borchers, Hans Joos,
Werner Theis, Klaus Helmers, Peter Stichel, Georg S\"{u}\ss mann, and Bert Schroer, because
we wanted to learn how to quantise the gravitational field. (Today I think it
must not be quantised (2007), because there are no observable effects of
quantised gravity: only the equations of state require quantised treatment). I
still remember the scene -- as though it happened yesterday -- when J\"{u}rgen
was the speaker of the day, and was cautioned by Helmers about an
inconsistency of his statement (on the supposedly unitary correspondences
between interacting and free fields): J\"{u}rgen understood immediately, was
upset, and threw the library copy of J \& R on the desk of our seminar room in
disgust, because the formal elegance of the book's presentation had fooled
him, had made him believe that the offered correspondence was mathematically
exact. This has been the only case of its kind which I can remember, in our
whole joint life.

Our scientific interests in General Relativity shifted gradually, from a
deeper understanding of its local structure towards a study of its global
properties: For instance, we could show how Newton's theory of gravity can be
obtained from Einstein's (more) general theory as an exact limit, by a
widening of its light cones (Trautman, 1964; my habilitation talk, 1965;
Ehlers, 1998; listening to J\"{u}rgen was always rewarding, even when he
returned to long-standing problems). Increasing insights were likewise gained
into the asymptotic behaviour of outgoing radiation, and also into topological
peculiarities -- like `trouser worlds' -- which are forbidden by the global
regularity of the spacetime geometry. And then there was the class of metrics
describing mass concentrations of extreme compactness, the Schwarzschild and
Kerr-Newman fields and their analytic continuations (through their `horizons')
which were eventually termed `black-hole' spacetimes by John Wheeler and Remo
Ruffini, in 1971; they will be the theme of the second section. Such non-local
problems were attacked originally by Roger Penrose, Brandon Carter, George
Ellis, and Stephen Hawking in Cambridge (England), later also by Bob Geroch
(Princeton), Werner Israel (Edmonton, Alberta), Kip Thorne (Caltech), Petr
Hajicek (Bern), and by a small number of others, including Seifert and Mueller
zum Hagen from our Hamburg group; (cf. my 1971 review at Halifax, published in
1972; also Ehlers et al, 1972, and Heusler, 1996). Exploiting Einstein's
theory kept us engaged.

Yet another branch of physics caught my attention when I returned from
Pittsburgh, in 1967, whilst J\"{u}rgen started his professorship at Austin:
Four young men wanted to start their diploma work with me, plus a fifth his
ph.d. work. We joined forces with my close friends Klaus Hasselmann and Gerd
Wibberenz, in an intense, wide-ranged seminar on statistical mechanics which
lasted successfully until 1971. Jordan's relativity seminar survived in
parallel, until his formal retirement in 1970, and even a few years beyond. In
addition, I was elected as the principal investigator of experiment 11 on the
German-American spaceprobe Helios, to test the validity of Einstein's theory
beyond the 1\% level -- a task that kept my collaborators and me actively
engaged from 1969 until 1979, passively even until 1983 -- even though our
experiment had to be sacrificed to the ten active ones on board when in both
missions, one of the two amplifier tubes burned out when data trasmission was
interrupted by range measurements.

All these activities allowed me to extend my insight gradually throughout most
of physics, at CERN (1972), Bielefeld (1973-1974.3), and Bonn (1978-now), with
several informative visits abroad in between. I moved from general relativity
through quantum field theory and statistical mechanics to cosmology --
after the detection of the 2.73 K background radiation, and after Rolf
Hagedorn's proposal of a highest temperature, near the pion rest energy, which
would have a distinct impact on the first nuclear reactions in the Universe --
and subsequently into astrophysics, planetary physics, geophysics, and even
biophysics, all of which fields have left traces in my 2005 Springer book
\textit{Astrophysics, A new Approach}, and in my 2008 article contributed to
the book \textit{Against the Tide}. They brought me increasingly into conflict
with general wisdom, through the attempt to keep physics as clean as possible
from inconsistencies, a demand that ought to be self-understood, but
unfortunately is not. Such conflicts are collected in the form of
79\ `alternatives' in my 1998 birthday book `Understanding Physics' -- the 
same book, and symposium, to which J\"{u}rgen contributed his `Newtonian 
limit' version (1998) -- and had
grown to 100 alternatives in the\ 2005 edition of the Springer book; their
number has meanwhile reached 124. The alternatives emerged through the desire
of thorough understanding, and whenever I had a chance to meet J\"{u}rgen, I
discussed some of them with him, even if and when he did not call himself an
expert in that particular area. To know J\"{u}rgen's reaction has always been
helpful. His co-authored book on gravitational lensing (1992) ranks high in
the world list of citations. Our last oral, long conversations took place in
the summer of 2006, during the 11th Marcel Grossmann meeting in Berlin, in
between the main lectures, but mainly at leisure, on banks in a nearby park.

You care to know what my alternatives are all about? The two toughest among
them,15 and 27, claim that we have not detected a single black hole in the sky
yet, no matter whether of stellar mass, or supermassive, and that we may never
detect any. But this is the subject of the next section. Most alternatives
reached me similarly to how they reached Fred Hoyle, according to his 1955
book \textit{Frontiers of Astronomy}: Certain explanations lack convictive
power and/or lack the harmony of the grand design, noticeable on the timescale
of seconds to minutes. Sure enough, with the relevant documents at hand, it
then does not take more than a couple of days until conjectures mature into
solid reasonings. Tom Gold was my great example and teacher in pursuing them,
during multiple
encounters in multiple places and on various occasions. I owe him a lot of
insight, as I owe to J\"{u}rgen.
So let me end this sketchy
history of 54 years of close interaction with J\"{u}rgen by adding three
further, rather recent examples.

The first example is my alternative 101: the second law of thermodynamics, its
continuum form: it cannot even be found under this name in Landau \& Lifshitz,
nor in Ehlers (1973). L\&L VI \S49 call it shily the `heat transport
equation'. The latter describes quantitatively the rate at which ordered
kinetic energy is converted into disordered one, and interpolates the discrete
entropy formulae of box thermodynamics, hence merits the name `entropy theorem',
and serves as a restriction to many
thermodynamic processes which would otherwise be permitted, under the sole
constraint of 4-momentum conservation (Kundt 2007). It has many applications
to various practical problems: How do the galaxies, stars, and planets
generate their large-scale magnetic moments? How do plants raise the needed
water to their crowns, against gravity? How do celestial objects accelerate
ions and/or electrons to huge particle energies, $\lesssim10^{20.5}$eV? The
literature contains answers to these 3 questions which violate the second law
(I claim). I was glad that during a short e-mail correspondence with J\"urgen,
I could settle this fundamental point with him.

The second example is my alternative 2,\ which appeared in print as a
(somewhat distorted) letter to Nature in 1976: What is the physical meaning of
the expression called `black-hole entropy' by Stephen Hawking in 1974? I
argued that his (gigantic) expression measures the entropy liberated during
the hole's evaporation, lasting 10$^{67}$years for a solar mass, not during
its formation, measuring in microseconds for the same mass (also: Kundt 2009).
J\"{u}rgen had no objections to my letter when I told him about it, whereas my
friendship with Stephen degraded.

The last example concerns a weird claim in the literature which reached me
shortly before J\"{u}rgen's 70th birthday (to whose celebration in Golm I was
invited). Abhas Mitra, an Indian physicist of by and large respectable career,
provoked the Western establishment via internet with the claim that the
black-hole literature was faulty, that BH spacetimes could not exist. He
bombarded me with calculations and quotations (via e-mail) which did not
convince me, but I could not point at an obvious error. I asked a number of
friends from the old days for help, with the same result. Finally, within less
than a week after his birthday symposium, J\"{u}rgen caught A.M.\ on an
oversight: he had confused partial derivatives with total ones. It needed
J\"{u}rgen's care and patience to find the mistake.

And this was by far not my last communication with J\"{u}rgen: our last
encounter was in July 2006, in Berlin; his last mail reached me in December
2007, containing the statement that Euclidean cosmologies -- as proposed
recently by a number of experts -- had not convinced him.

\section{ A short history of Black Holes}

This second section will be devoted to the role of black holes in present-day
astrophysics, my alternatives 15 and 27, (Kundt 2005). What are the black-hole
spacetimes? Why were they considered relevant for astrophysics? How were the
black holes hoped to be discovered? And why do I question all those hundreds
of claimed detections?

As sketched in section 1, exact solutions of Einstein's field equations have
been searched for ever since their proposal in 1915, but the systematic
finding of large classes with specified properties had to wait for the 50s and
60s. Among the first goals were solutions for isolated massive bodies, both
without and with spin, i.e. static and stationary. The first (1-parametric)
class was found almost immediately, by Karl Schwarzschild, in 1916. The second
(2-parametric) class was first published in 1963, by Roy Kerr; it required his
skill and perseverance. The fascinating story of its 8-yr-long discovery is
described by him in his 2006 plenary talk in Berlin, printed in 2008. See also
Heusler (1996), and Dautcourt (2009), for supporting descriptions.

The primary goal of the community was to find the exact outer vacuum solutions
for compact, massive bodies. During years of hard work, it came as a surprise
that the complete set of such spacetimes was only 3-parametric, interpreted as
the mass $M$, angular momentum $J$, and charge $Q$ of the field-generating
body; John Wheeler coined it by ''a black hole has no hair''. Another surprise
was that whereas the outer vacuum fields were stationary or even static, their
analytic continuations towards their center changed into time-dependent
geometries, across a lightlike (or null) bounding hypersurface termed
`horizon'. This structure of the complete class of black-hole spacetimes -- so
termed in 1971 by both Remo Ruffini \& John A.Wheeler\ and by Yakov Zel'dovic
\& I.D.Novikov\ -- was interpreted as the vacuum spacetime geometry left
behind a collapsing body in both its outside, and inside world. Less
symmetric, collapsing bodies would radiate away all their non-fitting higher
multipole moments, it was thought -- and calculated by R.H.Price (1972) --
before they cross their horizon, and approach a black-hole geometry at spatial
and lightlike infinity.

Once the hairlessness of black holes is ultimately proven, their multipole
moments are those of the (stationary) Kerr-Newman class of spacetimes, which
are
\begin{equation}
M_{n}=M\text{ }a^{n}\text{ },\text{ }Q_{n}=Q\text{ }a^{n}\text{ },\text{
\ }n\geq0\label{multipoles}%
\end{equation}
according to Newman \& Janis (1965), in which \ $ac:=$ $J/M$ \ is the hole's
specific angular momentum, $M_{n}$ are the hole's \{mass, spin\} multipole
moments for \{even, odd\} n, and $Q_{n}$ are correspondingly the hole's
\{electric, magnetic\}moments. As already stated, all initial deviations of a
collapsing body's higher multipole moments from the above sequences are
thought to be radiated to infinity, via gravitational and electromagnetic
waves. Remarkably, a charged, spinning black hole has the same gyro-magnetic
ratio $Q_{1}/M_{1}=Q/M$ as the electron.

The 3 parameters $M,J,Q$ cannot all take arbitrary values: The area $A$ of the
(outer) event horizon of a rotating, charged black hole (BH) reads
\begin{equation}
A/4\pi=2m^{2}[1+\sqrt{1-(a/m)^{2}-(q/m)^{2}}-(q/m)^{2}/2]\label{horizon}%
\end{equation}
with: $m:=GM/c^{2},\text{ }q:=\sqrt{G}Q/c^{2}$;
where $m$ is called the Schwarzschild length of the BH, and the
length $q$ measures its electric charge, in natural units. This area $A$
should be positive and real, forbidding superluminal rotation, and excessive
charging. It cannot shrink during accretion, hence the inference that energy
can be extracted from a spinning BH by braking its rotation -- whereby its
mass may even shrink\ -- but only to the extent that this mass stays above its
`irreducible' value $M_{irred}:=M$ $\sqrt{1-(a/m)^{2}-(q/m)^{2}}$. It was
speculated that such an extraction of energy could happen via accretion of
matter, part of which is forcefully re-ejected from the ergosphere -- \ with
excess forward angular momentum \ -- whilst a less energetic fraction falls
in, and brakes the hole. In the case of two merging BHs, their horizon areas
$A_{j}$ should satisfy the inequality:\ $A_{3}\geq A_{1}+A_{2}$ , and again it
should be possible to extract energy, with a theoretical efficiency
$\varepsilon$\ given by $\varepsilon$ = $(m_{1}+m_{2}-m_{3})/(m_{1}+m_{2})$
$<$%
$1/2$. Instead, realistic situations -- if such exist -- may have efficiencies
which are lower by several orders of magnitude, depending on the state
parameters (density, opacity) of the accreted material, and on its advected
angular momentum (Kundt 2009).

The horizon area $A$ serves to define the (famous) Schwarzschild radius
$R_{S}$ of a BH via
\begin{equation}
R_{S}:=\sqrt{A/4\pi}=2GM/c^{2}=3\text{ Km }(M/M_{\odot})\text{ \ ,}%
\label{radius}%
\end{equation}
the latter two equalities for vanishing spin and charge; no self-gravitating
object can have a smaller size than given by $R_{S}$. The corresponding
critical mass density $\rho_{crit}$ for BH formation follows as $\rho
_{crit}=\rho_{nucl}$ $(7M_{\odot}/M)^{2}$; it exceeds nuclear density for a
solar mass, but requires only a density of a terrestrial high vacuum for
galactic masses.

All these highly structured and aesthetic results about the BH solutions of
Einstein's gravitational equations -- obtained essentially during the 60s\ --
meant a great challenge to the minds of the young astrophysics community, to
search for their verification in our Milky Way and beyond; where do the most
convincing candidates hide? When the first pulsar was detected, in 1967 by
Jocelyn Bell, it was soon interpreted (by Tom Gold) as a neutron star, with a
mass in the vicinity of $1.4M_{\odot}$. Could neutron-star masses have a broad
distribution, with an upper cutoff near their stability limit, some
$3M_{\odot}$, and BHs beyond? Can BHs be born directly in supernova
explosions, of the most massive stars? Or can they form via neutron-star
accretion of mass from a binary companion, in X-ray binaries, (of which the
first pulsating one was again detected by Jocelyn Bell-Burnell, in 1974)?
Either of these two possibilities sounded plausible to us, including myself
when I heard, spoke, and wrote about them in 1972, 1973, and 1976. Most
impressive for me was the Varenna summer School in 1972 where the newly
suspected X-ray binary system Cyg X-1 was discussed at length by the high-brow
assembly, even during the lunch break, on the beach of lake Como, as the
best-bet BH candidate. And there was another class of BH candidates, typically
a million times larger in mass, often highly variable, located at the centers
of galaxies -- the supermassive black holes (SBHs)\ -- whose power can exceed
that of their host galaxies by a factor of up to $10^{2}$. Its nearest
representative, Sgr A*, at the rotation center of our Galaxy, received its
name only in 1982 (by Brown), but was already considered an SBH candidate in
1971, by Donald Lynden-Bell and Martin Rees. Only at the 1976 Texas Symposium
(at Chicago), the central engines (CEs) of all the active galactic nuclei
(AGN) were irreversibly called BHs by Martin, after some five years of
friendly competition between them and `supermassive stars' (SMSs), `magnetoids',
and `spinars'.

All these thoughts and facts were more than tempting: Einstein's GR had
finally found its true crowning, by offering the likely structure of many of
the most luminous sources in the sky. Whenever a physical theory had made a
convincing and testable prediction in the past -- I had learned\ -- this
prediction had soon been verified. I still needed friends like Hans
Heintzmann, to introduce me to real astrophysics, and Tom Gold who told me
about Jearl Walker's `flying circus' of physics, to form a judgement of my own.

In the present case, conclusive tests were not easy to achieve: no expected BH
was near enough to Earth to just go there and look at it. Nature could have
put up hurdles to their formation. None of the proposed formation mechanisms
was ultimately conclusive. Angular-momentum conservation and explosive nuclear
burning could delay their formation, or even prevent it for aeons. Besides,
most of the suspected BH sources were (i) extremely powerful, (ii) extremely
variable, and (iii) spectrally hard. The efficiency $\varepsilon$ of BH
engines in converting accreted mass-energy into (hard) radiation was (i) often
judged very high, of order 10\% and higher, but never realistically assessed:
ordered gravitational infall into large BHs could be almost traceless for the
outside world, meaning $\varepsilon\ll1$. Next, high variability of a source
(ii) requires strong deviations from axial symmetry, which (spinning)
neutron-star sources are thought to achieve via strong transverse magnetic
moments. BHs cannot anchor a transverse magnetic moment, was the discouraging
conclusion of my younger collaborators King and Lasota, in 1975; their
variabilities would have to be blamed on their accretion disks, a difficult
task to perform. And the spectra of the suspected BH sources, which (iii) got
harder and harder thoughout the years, with improving observational
facilities, have now grown into the TeV range (instead of peaking at X-ray
energies); don't they require rapidly rotating strong magnetospheres for their
boosting, according to the estimate: $\Delta W=e\int(\overset{\rightarrow
}{E}+\overset{\rightarrow}{\beta}\times\overset{\rightarrow}{B})$
$d\overset{\rightarrow}{x}$ $=10^{21}$ eV $(\beta_{\bot}\times
B)_{12}(\Delta x)_{6.5}$, (2005)? After an unsuccessful (1 yr) bet with Ed van
den Heuvel fixed in 1976 \ -- \ in which he mentioned a possible neutron star
(n*) in the Cyg X-1 system (!) \ -- \ I convinced myself that an n*
interpretation was indeed plausible (1979).

Returning to the possible formation modes of BHs, my decades-long engagement
in supernova (SN) explosions led me to the conviction that core collapse is
the only viable SN mechanism: Some day, the burnt-out, magnetized core of a
massive star, of (Chandrasekhar) mass 1.4 $M_{\odot}$, collapses under its own
gravity, transfers a significant fraction of its spin energy to its overlying
magnetosphere, by flux winding, and ejects its extended, overlying, massive
envelope by the joint action of a magnetic torque plus an adiabatic expansion
of the magnetosphere's decay product, a relativistic cavity (2008c). This
mechanism is robust, sufficiently energetic to leave (even) a ms pulsar
behind, and satisfies all the (phenomenological and stability) SN constraints.
It may well be successful up to the highest stellar masses. It may therefore
be extremely difficult to form a stellar-mass BH right away, in a SN
explosion. No convincing case is known to me, after several decades of
checking the literature, in which a SN had given birth to a BH: Neutron stars
can easily hide at the centers of SNe, and more and more of them have been detected.

So the easiest way to make a stellar-mass BH appears to be to dump matter on a
n*. Now there is the Eddington limit, which prevents a n* from accreting mass
at a rate exceeding $10^{-8}M_{\odot}/$yr, due to the Leidenfrost phenomenon.
All excess matter, supplied by a binary companion star, will accumulate in a
surrounding (accretion) disk, conceivably piling up to five or more solar
masses during its lifetime. Such massive, self-gravitating disks have the
reputation of being unstable, for no ultimate reason (1979, 2005). They make
the encircled n* look superheavy. Binary systems of this kind are known as
black-hole candidates (BHCs). Almost all other properties of the BHCs are
indistinguishable from neutron-star sources: their hard spectra, fast
variabilities, QPOs, lightcurve anomalies and strong emission lines,
superhumps during outbursts, and in particular their jet-forming capabilities
(Kundt \& Fischer 1989, Kundt \& Krishna 2004).

What about the SBHs, with their relativistically broad iron lines in emission,
which are supposed to lurk at the centers of all massive galaxies? Have they
formed, and grown, during cosmic epochs? Fig.1 shows a representative set of
galactic rotation curves $v(r)$, or rather of their (equivalent) average
surface-mass densities $\sigma(r)$ $\sim v(r)^{2}/r$ . Its upper right grey
area is the BH corner; it is free of entries. Mass distributions outside of
the BH corner are stabilized against collapse: radially by centrifugal forces,
and vertically by pressure forces. No single CE in the plot is unstable
towards gravitational collapse. Instability towards collapse is expected for
masses in excess of $10^{10.5}M_{\odot}$, implying that galactic rotation
speeds reach the speed of light at their periphery.%

\begin{figure*}
\noindent\includegraphics[width=\textwidth]{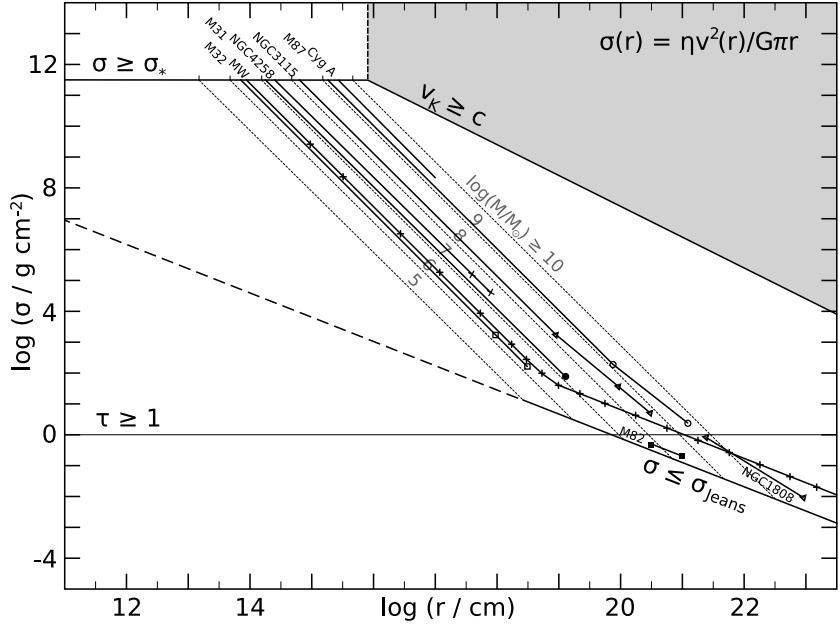}
\caption{\label{fig:rot}
Complete rotation curves  -  with $10^{11} \lesssim r/$cm $ \lesssim
10^{23.5}$  -  for a representative set of well-sampled
galaxies, taken from (Kundt 2008a). For a better understanding of galactic centers, the ordinate
presents average surface-mass density  $\sigma(r) \lesssim v^2(r)/G \pi
r$ (instead of rotational
velocity $v(r)$): Whilst $\sigma
(r)$ is tiny in the outer parts of a galaxy, where it is controlled by
Jeans instability (to star formation), it grows considerably with decreasing $r$, but cannot exceed
stellar values ($\sigma_{*} \approx10^{11.5}%
$ g/cm$^2$), due to pressure forces, hence sets a
bound on revolution speeds near the center. Observations indicate that galaxies have ringlike
domains of insignificant (gravitating) mass density, between $\gtrsim
10^{14}$cm and $\lesssim10^{20}$cm,
in which their rotation is solely controlled by the mass of their central engine (CE), and $M(r) = const$.
Note that the detected CE masses all stay below the BH formation limit of $10^{10.5}%
M_{\odot}$  -
marked in grey  -  beyond which they would enforce (among others) extremely relativistic galactic
revolution speeds.}
\end{figure*}%

When I started contemplating about the nature of the CEs of active galaxies,
back in 1977, it occurred to me that the earlier pursued models mentioned
above, SMS, magnetoid, and spinar, already caught essential features of the
central engines, but had predicted properties which slightly disagreed with
the facts, like evolving (spin) periods, and short lifetimes. Instead, gaseous
galactic disks have predicted mass spiral-in rates of order $M_{\odot}/$yr ,
they will continually re-charge whatever sits at their center. Why then not
simply study the innermost parts of galactic disks, whose mass densities per
area are expected to grow in proportion to $r^{-1}$, and approach stellar
values in their solar-system sized centers? These centers should behave as
flat stars, or as (nuclear-) burning disks (BDs). Their radiative efficiencies
should (i) easily exceed those of BH accretors; they should have (ii) strongly
variable outputs, and (iii) generate relativistic pair plasma in their reconnecting
magnetized coronae, ready to drive jets; their burning (iv) produces 
large quantities of nuclear ashes, in particular of iron (Turnshek
1988), and they should certainly (v) blow strong winds (and thereby discharge 
in mass). And moreover,
(vi) these BDs would conform with the cosmic evolution of the AGN
phenomenon plotted in fig.2: Their masses would shrink with age -- not grow
with age, as BH masses must \ -- when they discharge efficiently, via their
strong winds. Such ideas found support by people like Peter Scheuer, John
Biretta, Ski Antonucci, and others, and were published, among others, in 2002,
2008a,b, and 2009. BH proponents are aware of the inconsistent (sign of)
evolution documented in fig.2, of course, but have forced it to agree
with\ their expectation by using words like ''downsizing'',
''antihierarchical'', ''co-evolution'', ''feedback from SNe and AGN'', and the
like; the BH paradigm must not be sacrificed.%

\begin{figure*}
\noindent\includegraphics[width=1\textwidth]{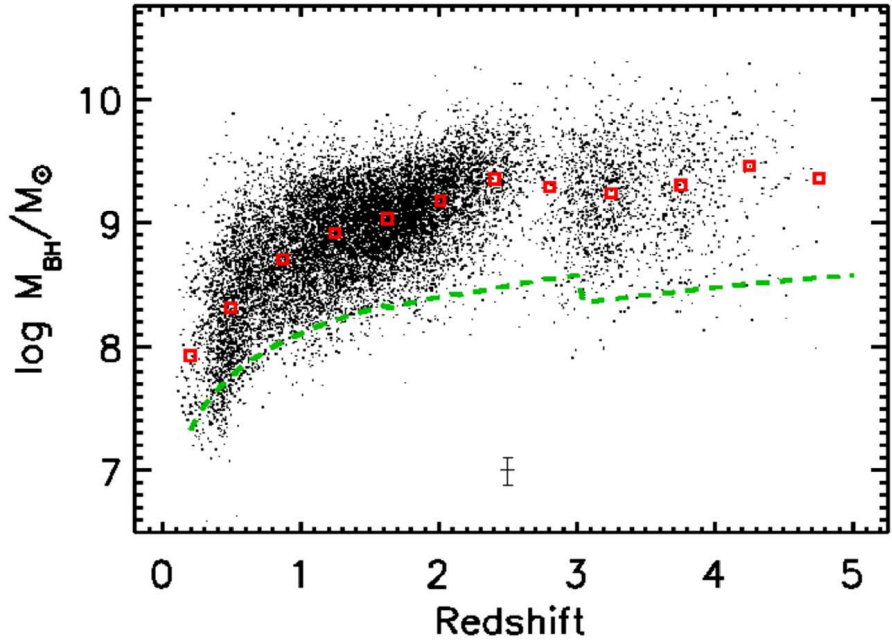}
\caption{\label{fig:dis}
(Estimated) mass distribution of 14,584 quasar central engines (CEs) with $z \geq
0.2$, as functions of redshift $z$,
from the Sloan Digital Sky Survey Data Release 3, within an effective sky area of 1644 deg$^2$, taken from
Vestergaard et al (2008). Squares denote median masses in each redshift bin. The dashed curve indicates
faint SDSS flux limits.}
\end{figure*}%

Two recent discoveries are worth mentioning. One of them is the QSO SDSS
J1536+0441 with its two broad-line emission systems (plus one unresolved
absorption line system), interpreted as a SBHC system by Boroson \& Lauer
(2009), as an exceptional case among $10^{4.24}$ similar entries in the
SDSSurvey. The binary BH interpretation is motivated by the two broad-line
systems in the spectrum, which differ in speed by $10^{3.5}$Km/s, and asks for
masses of $10^{8.9}$ and $10^{7.3}M_{\odot}$. As you may expect, my preferred
interpretation of this source is a single BD with a second broad-line region
(BLR), outside of the first (and normal) BLR, which stems from the rare event
of a somewhat faster ejection of iron traversing a galactic hydrogen cloud.

The second discovery concerns SgrA*, the CE of our Galaxy, on which I wrote a
detailed review in 1990, explaining all known
observations within the BD model. (This paper was submitted before a
compelling mass determination of SgrA*; I was not the only one, then, who got
it wrong, by a factor of $\gtrsim10^{3}$). Another review on this subject was
written by my younger friends Fulvio Melia and Heino Falcke, in 2001. They
kindly mention my 1990 model in their section 5.3 called ''Alternatives to the
Black Hole Paradigm'', but state that ''it can be comfortably ruled out, based
on the low NIR flux at the position of SgrA* ''. They did not ask me for my
opinion, before submission, on their devastating statement, perhaps because
they were afraid of getting involved in an embarrassing dialogue on their
draft. I only saw it years after publication, by accident, whilst working on
several unrelated problems. Had I been asked, I would have reminded them that
the underluminosity of SgrA* at NIR frequencies is a problem which their BH
model shares with mine: An engine that emits a broadband spectrum between high
radio frequencies and TeV energies, with a power of $\gtrsim10^{41}$erg/s in
its (inferred) relativistic wind, and $\gtrsim10^{39}$erg/s in its (mapped)
thermal wind, should not be fainter than (the visible) bright stars at
wavelengths above $2\mu$m (where foreground absorptions disappear), unless it
is screened by a thick, dusty absorption layer. I would also have reminded
them that from Earth, we look at SgrA* through our Galactic disk, like for
(extreme) Seyfert galaxies of type 2 whose hot centers are thought to be
occulted by dusty tori. These ''tori'' have never been resolved completely,
but are thought to extend radially to distances of $\lesssim1$ pc, and extend
to fairly high ($\gtrsim30^{0}$) latitudes. The tori may well be dusty winds
from the hot central disk, remotely similar to the winds from supergiant
stars, which are known to be opaque at almost all wavelengths. For a central
galactic disk, the windzone may well have the shape of a cusp, envelopped by a
dusty collar that occults the hot core from view at low angles. In the case of
SgrA*, the strong, blue- and red-shifted Br$\gamma$ and Br$\alpha$ lines
visible in fig.7 of my 1990 review, at distances of $\lesssim1$ lyr from the
center, are witnesses of a hot, screened interior.

Whilst talking about SgrA*, the nearest among all supermassive CEs in the
Universe and therefore the best supermassive BHC of all, I should repeat what
I already reported in my 2009 Portobello lecture notes: its non-BH character
is now manifest to me in five independent ways. They are: 1) The 16yr-Kepler
ellipse of star S2 around it has been reported by Frank Eisenhauer (in
November 2007)\ not to close, by 3$^{0}$. 2) The mass estimate of SgrA* has
grown throughout the past years -- during its approach\ -- from $10^{6.41}$to
$10^{6.58}M_{\odot}$. 3) The distance estimate of SgrA* implied by the
monitored test-star orbits has reached 8.33 Kpc, whilst other distance
estimates claim $\lesssim$ 8.0 Kpc. 4) The gigantic wind from SgrA*, mapped by
the blown-off windzone tails from $\gtrsim8$ nearby stars, at distances of
$\lesssim1$ lyr, and in the light of the redshifted Br$\alpha$ and blueshifted
Br$\gamma$ line, of mass rate $10^{-2.5}M_{\odot}$/yr, and speed
$\lesssim10^{3}$Km/s,\ exceeds what can be achieved by a collection of
surrounding stars, and much more so what can be achieved by a starving BH. And 5) the mysterious brightening (by 0.5 mag), and angular offset of S2 near periastron, in 2002, asks for a dense local environment (Gillesson et al 2009).

With this ends my short history of Black Holes, as of 2009. I once helped
studying them, and advertised them in 1972, and in my two semi-popular reports
in 1973 and 1976. During that latter year, first doubts in their reality
emerged, and made me wonder if not more conservative sources could yield
better descriptions of the proposed candidates. And by 1979, I was no longer
sure of their presence, anywhere among the vast zoo of astrophysical sources.
With this I am not saying that black holes were unphysical: they may have
formed somewhere, or may form somewhere in the future, when a lot of hydrogen
has been burnt into iron. Without nuclear power, BH formation is difficult to
prevent. But we live in a young Universe, and I expect BH sources to be
rather inconspicuous compared with non-BH sources: with low power, low
variability, and soft spectra (below $\gamma$-rays). Our life on Earth should
be hardly influenced by them. Does this sound like bad news? I do not think
so: our planet Earth offers an overabundant richness of physics problems, even
without BHs. Let us enjoy all of them!

\bigskip

\begin{acknowledgements}
My cordial thanks go to Ole Marggraf, for help with the electronic
data handling, to Tim Eastman for repeated help and encouragement, and to Gernot Thuma for the manuscript.
\end{acknowledgements}

\end{document}